\documentclass[conference,a4paper]{IEEEtran}
\IEEEoverridecommandlockouts
\usepackage{cite}
\usepackage{amsmath,amssymb,amsfonts}
\usepackage{algorithmic}
\usepackage{graphicx}
\usepackage{textcomp}
\usepackage{xcolor}
\def\BibTeX{{\rm B\kern-.05em{\sc i\kern-.025em b}\kern-.08em
    T\kern-.1667em\lower.7ex\hbox{E}\kern-.125emX}}

\allowdisplaybreaks
\usepackage[acronym,nogroupskip,nomain,nonumberlist,style=alttree,section=subsection]{glossaries}

\setglossarystyle{alttree}
\newacronym{lfm}{LFM}{Local Flexibility Market}
\newacronym{dam}{DAM}{Day-Ahead Market}
\newacronym{afrr}{aFRR}{automatic Frequency Restoration Reserve}
\newacronym{fl}{FL}{Flexible Load}
\newacronym{dg}{DG}{Distributed Generation}
\newacronym{bess}{BESS}{Battery Energy Storage System}
\newacronym{soc}{SOC}{State of Charge}
\newacronym{ev}{EV}{Electric Vehicle}
\newacronym{hvac}{HVAC}{Heating, Ventilation and Air Conditioning}
\newacronym{der}{DER}{Distributed Energy Resource}
\newacronym{mo}{MO}{Market Operator}
\newacronym{tn}{TN}{Transmission Network}
\newacronym{dn}{DN}{Distribution Network}
\newacronym{vpp}{VPP}{Virtual Power Plant}
\newacronym{ems}{EMS}{Energy Management System}
\newacronym{eoa}{EOA}{Energy Optimization Assessment}
\newacronym[\glslongpluralkey={Local Energy Communities}]{lec}{LEC}{Local Energy Community}
\newacronym{dc}{DtC}{Data Centre} 
\newacronym{ai}{AI}{Artificial Intelligence} 
\newacronym{milp}{MILP}{Mixed Integer Linear Programming} 
\newacronym{qos}{QoS}{Quality of Service}

\begin{document}

\title{Exploiting Data Centres and Local Energy Communities Synergies for Market Participation
\thanks{This work was supported by the FPU grant (FPU19/03791) founded by the Spanish Ministry of Education, by Ministerio de Ciencia e Innovación through projects TED2021-132339B-C42, by Horizon Europe Programme through projects HORIZON-CL5-2022D3-01 Ref: 101096787, HORIZON-CL5-2022-D4-02-04 Ref: 101123556 and, by the University of Málaga.}
}
\author{
\IEEEauthorblockN{
Ángel Paredes\IEEEauthorrefmark{1},
Yihong Zhou\IEEEauthorrefmark{2},
Chaimaa Essayeh\IEEEauthorrefmark{3},
José A. Aguado\IEEEauthorrefmark{1},
and Thomas Morstyn\IEEEauthorrefmark{4}}
\\
\vspace{-3mm}
\IEEEauthorblockA{\IEEEauthorrefmark{1}Department of Electrical Engineering, University of M\'alaga, Spain, \{angelparedes, jaguado\}@uma.es}
\IEEEauthorblockA{\IEEEauthorrefmark{2}School of Engineering, The University of Edinburgh, U.K., yihong.zhou@ed.ac.uk}
\IEEEauthorblockA{\IEEEauthorrefmark{3}Department of Engineering, Nottingham Trent University, U.K., chaimaa.essayeh@ntu.ac.uk}
\IEEEauthorblockA{\IEEEauthorrefmark{4}Department of Engineering Science, University of Oxford, U.K., thomas.morstyn@eng.ox.ac.uk}
\vspace{-6mm}
}


\maketitle

\begin{abstract}
The evolving energy landscape has propelled energy communities to the forefront of modern energy management. 
However, existing research has yet to explore the potential synergies between data centres and energy communities, necessitating an assessment on their collective capabilities for cost efficiency, waste heat optimisation, and market participation.
This paper presents a mixed integer linear programming model to assess the collaborative performance of energy communities, data centres and energy markets.
The evaluation focuses on the efficient use of waste heat and the flexibility of job scheduling while minimising system energy costs and maintaining quality of service requirements for data centres.
Our results, based on realistic profiles of an energy community and a data centre, showcase significant benefits of these synergies, with a 38\% reduction in operating costs and an 87\% decrease in heat demand.
\end{abstract}


\section{Introduction}

\IEEEPARstart{T}{he} energy landscape is undergoing a major transformation, marked by an increasing emphasis on sustainability and renewable energy sources.
Energy communities are gaining popularity as a means of enabling local members to pool resources and collectively manage energy production and consumption \cite{F.G.Reis2021}. 
At the same time, and as we move into the future, the growing use of artificial intelligence and other computing services is leading to a rising demand for enhanced computing capabilities \cite{Guo2021}.
The synergies between renewable energy, efficient computing capabilities, and the emergence of \glspl{lec} are of paramount importance in addressing the evolving energy needs of our society.


In the realm of \gls{dc} resource optimisation, authors in \cite{Wang2019a} underscore the importance of optimising resource consumption, drawing from a comprehensive five-year study of the Titan System. 
Reference \cite{Yin2023} explores Internet \glspl{dc}' potential as energy prosumers, emphasising their flexibility in integrated electricity-heat systems, contributing to enhanced energy efficiency and renewable energy integration. 
Addressing thermal challenges, one research effort \cite{Garimella2013} recognises the significance of thermal management in the IT industry's pursuit of sustainable practices within \glspl{dc}. 
Additionally, an innovative approach \cite{ZacharyWoodruff2014} envisions \glspl{dc} as distributed nodes providing waste heat for other building heating needs, thereby reducing overall energy consumption, showcasing adaptability and resourcefulness. 
Furthermore, a study \cite{Borland2023} delves into the application of waste heat from \glspl{dc} in vertical farming, highlighting opportunities to reuse waste heat, aligning with the trend toward resource optimisation and sustainable practices in the \gls{dc} industry. 

In light of these developments, \glspl{lec} play a pivotal role in the energy landscape, with potential for leveraging \glspl{dc}' waste heat and flexibility in energy markets. 
Notably, the need for innovative sources of heating and cooling within \glspl{lec} is a recurring theme, as highlighted by \cite{Fouladvand2022}. 
This study underscores the necessity for further exploration of innovative community-based initiatives. 
Besides, the importance of waste heat recovery and its integration with renewable energy sources to achieve sustainability goals has been consistently emphasised in research \cite{Huang2020}. 
These findings collectively highlight the pressing requirement for \glspl{eoa} and standardised energy metrics to ensure the long-term sustainability and efficiency of modern energy systems. 
While innovative \gls{eoa} have been proposed to potentially reduce costs and enhance grid flexibility \cite{Srithapon2023}, there is a lack of evidence regarding the contribution of \glspl{dc} to this endeavour when integrated with \glspl{lec}. 

The synergies between \glspl{dc} and \glspl{lec} has been analysed through coordinated building heat systems and electric vehicle charging \cite{Backe2021} and demand response programs \cite{Zhang2022}, showing huge potential for efficiency increase.  
In addition, \glspl{dc} have the potential of providing market services without jeopardising their \gls{qos} requirements using approaches such as dynamic voltage and frequency scaling, job preemption, and hardware power capping \cite{Jahanshahi2022}. 
Nevertheless, despite previous efforts, there is a research gap regarding how \glspl{lec}, \glspl{dc}, and energy markets interact and benefit each other. Existing research mostly looks at them separately, missing out on potential collaboration benefits.

The main contribution of this paper is an evaluation of the value of integrating \glspl{lec} and \glspl{dc} for market participation focusing on the waste heat and job scheduling flexibility. 
This paper proposes a \gls{milp} model for this assessment which minimise the total energy cost. \gls{dc}'s \gls{qos} requirements are included using Big-M method. 

The rest of the paper is organised as follows. Section II describes the  \gls{lec} and \gls{dc} model. Then, Section III presents a Case Study based on realistic datasets for energy patterns and jobs duration. Lastly, Section IV concludes the paper. 

\section{Problem formulation}

\begin{figure}[b]
    \centering
    \includegraphics[width=\columnwidth]{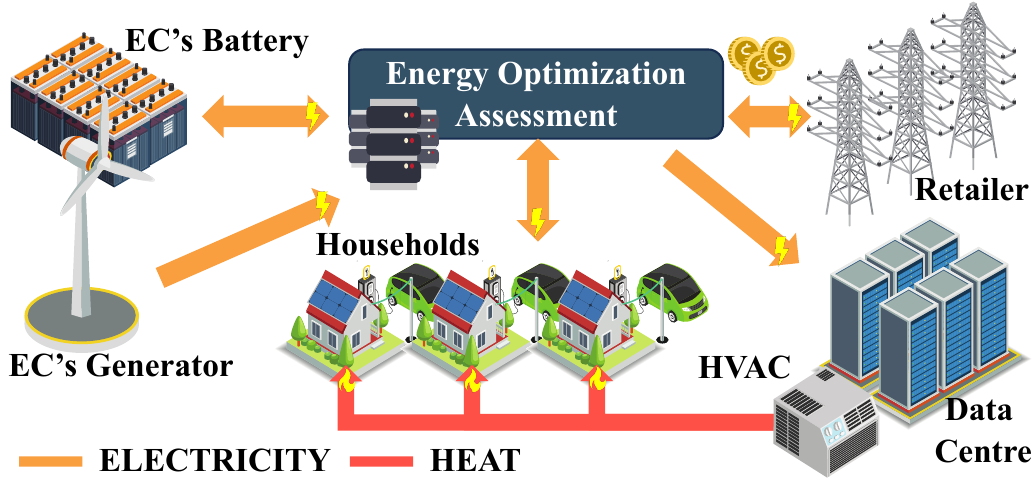}
    \caption{General overview of the proposed EOA module for the assessment of the synergies of LECs, DtCs and Energy Markets.}
    \label{fig:ec_model}
\end{figure}

The \gls{lec} comprises households with photovoltaic panels, flexible electricity consumption, and \glspl{ev}.
A battery and a small wind generator are the shared \gls{lec}'s resources. For heating, a shared \gls{hvac} system is employed, while private \gls{hvac} systems provide cooling. 
The synergies between \glspl{dc} and \glspl{lec} can be investigated using an \gls{eoa} module which oversees their joint energy and heat needs.
This integration is especially promising for urban scale with industrial technological hubs, given regulatory distance limits that allow connections to \gls{lec} shared resources.
Fig. \ref{fig:ec_model} illustrates the \gls{eoa} module for \glspl{lec} and \glspl{dc}. 

\subsection{Data Centre model}
In this section, we present a high-level model of the \gls{dc}, focusing on total computing power and \gls{qos} requirements.
The \gls{dc} operates with a baseline power determined by the job workload, denoted as $P_t^{wkl}$.
To enhance flexibility, the \gls{dc} can temporarily pause certain jobs and resume them during subsequent time periods \cite{Verma2015}. 
Let $p_t^{SB}$ be the running power of the jobs paused at time $t$ over a set of periods $\Omega_t$, $p_t^{RE}$ is the resuming power, $\mu_t$ is the mean duration of the jobs running, and $z_t^{DtC}$ is a binary variable indicating the pausing action.
The extra consumption resulting from the auxiliary operations to pause and resume the jobs is modelled by a positive constant $K^S > 1$.
\begin{subequations}
\begin{flalign}
    & p_t^{e,d} = P_t^{wkl} - p_t^{SB} + p_t^{RE} K^S & \forall t & \label{eq:power_datacentre} \\
    & 0 \leq p_t^{e,d} \leq P^{DtC} & \forall t & \label{eq:max_dtc} \\
    & 0 \leq p_t^{SB} \leq P_t^{wkl} z_t^{DtC} & \forall t & \label{eq:max_standby} \\
    & 0 \leq p_t^{RE} \leq P^{DtC} (1-z_t^{DtC}) & \forall t & \label{eq:max_resumming} \\
    & d_{t,t'} = \begin{cases} 
                    \frac{t'-t}{\mu_t} & \text{if } \sum_{\tau = 0}^{t'} P_{\tau}^{RE} \leq \sum_{\tau = 0}^{t} P_{\tau}^{SB} \\
                    0 & \text{otherwise}
                \end{cases} \hspace{-100pt} & \forall t, t'>t &  \label{eq:duration} \\
    & d_{t,t'} \leq K^{DELAY} & \forall t, t'>t & \label{eq:max_duration} \\
    & \sum_{\tau = 0}^{t} (p_{\tau}^{RE} - p_{\tau}^{SB}) \leq 0 & \forall t & \label{eq:re_sb} \\
    & \sum_{\tau = 0}^{|\Omega_t|} (p_{\tau}^{RE} - p_{\tau}^{SB}) = 0 & & \label{eq:finishing} \\
    & q_t^{DtC} = C  p_t^{e,d} + C^{MIN} & \forall t & \label{eq:heat_data centre}
\end{flalign}
\end{subequations}

The \gls{dc} demand $p_t^{e,d}$ is modelled by \eqref{eq:power_datacentre}, which must not exceed the \gls{dc} rating $P^{DtC}$ as \eqref{eq:max_dtc} states.
Power being paused $p_t^{SB}$ cannot be higher than the workload $P_t^{wkl}$ at time $t$, while the re-starting power $p_t^{RE}$ cannot be higher than the capacity of the \gls{dc} $P^{DtC}$, as \eqref{eq:max_standby} and \eqref{eq:max_resumming} respectively enforce.
The variable $d_{t,t'}$, computed in \eqref{eq:duration}, signifies the cumulative delay experienced by jobs that are paused at time $t$ up to $t'$.
It serves as a cumulative delay counter for the jobs paused at time $t$ until the resumption of the job.
However, this constraint cannot be directly solved by commercial solvers, needing a reformulation using the Big-M method, as detailed in Annex A.
$K^{DELAY}$ is defined as the maximum allowed delay to ensure \gls{qos} requirements in \eqref{eq:max_duration}. 
Equation \eqref{eq:re_sb} ensures that the \gls{dc} must have previously experienced a power reduction before power is being resumed, and \eqref{eq:finishing} enforce that all the paused power must be resumed to completion within the optimisation horizon. 
The heat generated by the \gls{dc} in its operation $q_t^{DtC}$ is modelled in \eqref{eq:heat_data centre} as a linear relationship $C$ with the power consumed $p_t^{e,d}$ plus a fixed term $C^{MIN}$ \cite{Yin2023}.

\begin{figure}[b]
    \centering
    \includegraphics[width=\columnwidth]{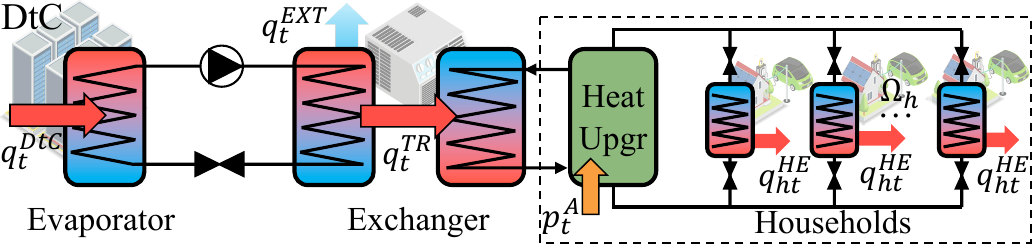}
    \caption{General overview of the Heat Recovery System. An evaporator draws heat $q_t^{DtC}$ from the \gls{dc}, while an exchanger and a posterior heat upgrading system transfer the heat $q_{h,t}^{HE}$ to the set of households $h \in \Omega_h$.}
    \label{fig:heat_transfer}
\end{figure}

The heat recovery system, illustrated in Fig. \ref{fig:heat_transfer}, use an evaporator to extract the heat from the \gls{dc}.
After that, the heat exchanger channels a portion of this energy to a heat upgrading system, denoted as $q_t^{TR}$ with efficiency $\eta^{TR}$, while releasing the surplus $q_t^{EXT}$ into the surroundings. 
Thus $q_t^{TR} = q_t^{DtC}\eta^{TR}$. 
This heat upgrading is crucial due to the low-grade nature of the heat produced by the \gls{dc}, which might fall short of meeting all the heating demands within the \gls{lec}. 
If that is the case,  the system employs a power input $p_t^{A}$ to generate supplemental heat for the \gls{lec} with efficiency $\eta^{A}$.

\subsection{Energy Community Model}
\begin{figure}
    \centering
    \includegraphics[width=\columnwidth]{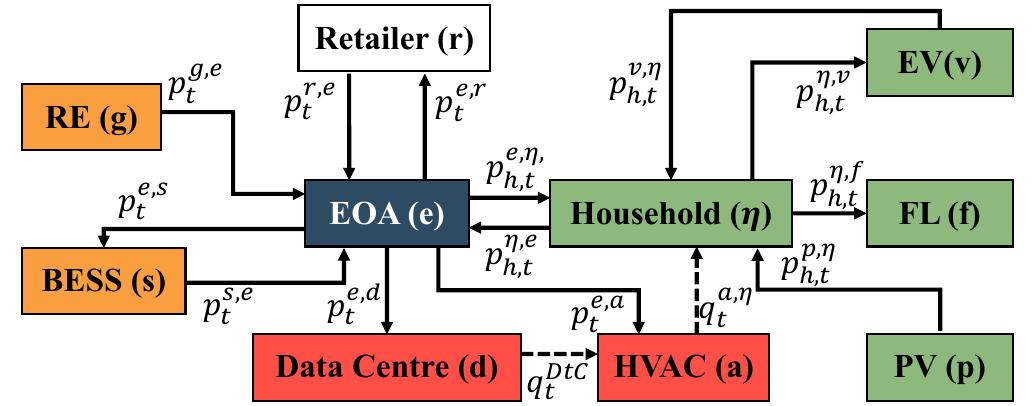}
    \caption{General overview of the proposed EOA for the joint operation of LECs and DtCs. Solid lines represent electric energy flows, while dashed lines represent heat transfer.}
    \label{fig:ems}
\end{figure}

The power flowing among the different elements is depicted in Fig. \ref{fig:ems}. 
Power can be bought from $p_t^{r,e}$ or sold to $p_t^{e,r}$ the retailer at prices $\lambda_t^{DA}$ and $\lambda_t^{PPA}$, respectively. 
\gls{dc} can also benefits from its participation in \gls{afrr} markets by providing upward reserves at price $\lambda_t^{aFRR}$ pausing its jobs.
The \gls{eoa} module receives power $p_t^{g,e}$, $p_t^{s,e}$, $p_{h,t}^{\eta,e}$ from the wind turbine, the storage and, the households $h \in \Omega_h$, and delivers power $p_t^{e,s}$, $p_t^{e,d}$, $p_{h,t}^{e,\eta}$ to the shared storage, the \gls{dc} and, the households, respectively.
Besides, each household optimises their flexible demand $p_{h,t}^{\eta,f}$ and their \glspl{ev} $p_{h,t}^{\eta,v}$ considering the self-generation $p_{h,t}^{g,\eta}$ and the injection of the vehicle-to-house $p_{h,t}^{v,\eta}$. 
For the sake of simplicity, \eqref{eq:ems_ec} seeks to minimise the global objective of the \gls{lec}, minimising the total cost of energy provision.

\begin{subequations}
\begin{flalign}
    & \min_{vars} \sum_t (\lambda_t^{DA} p_t^{r,e} - \lambda_t^{PPA}p_t^{e,r} - \lambda_t^{aFRR}p_t^{SB}) & \label{eq:ems_obj}
\end{flalign}
Subject to,
\begin{flalign}
    \nonumber & p_t^{g,e} + p_t^{s,e} + p_t^{r,e} +  \sum_h p_{h,t}^{\eta,e} = p_t^{e,r} + p_t^{e,s} + p_t^{e,d} + \sum_h p_{h,t}^{e,\eta} \hspace{-100pt} & & \\ 
    & \hspace{130pt}   \hspace{-100pt} & \forall t & \label{eq:ems_balance} \\
    & p_{h,t}^{\eta,e} + p_{h,t}^{\eta,v} + p_{h,t}^{\eta,f} = p_{h,t}^{e,\eta} + p_{h,t}^{v,\eta} + p_{h,t}^{p,\eta} \hspace{-100pt} & \forall t, \forall h & \label{eq:household_balance} \\
    & p_{h,t}^{\eta,f} = P_{h,t}^{sch} + p_{h,t}^{u} - p_{h,t}^{d} & \forall h, \forall t & \label{eq:ec_fl_consumption} \\
    & \sum_t (p_{h,t}^{u} - p_{h,t}^{d}) = 0 & \forall h & \label{eq:ec_fl_total} \\
    & p_{h,t}^{\eta,v} \leq P_h^{EV} z_{h,t}^{ev} \quad p_{h,t}^{v,\eta} \leq P_h^{EV} (1 - z_{h,t}^{ev}) \hspace{-100pt} & \forall h, \forall t & \label{eq:ev_ch_dis} \\
    \nonumber & soc_{h,t} = soc_{h,t-1} + (p_{h,t}^{\eta,v} \eta_{h,t}^{CH} - p_{h,t}^{v,\eta}/\eta_{h,t}^{DIS}) \Delta t  \hspace{-100pt} & & \\
    & \hspace{120pt} - P_{h,t}^{DRIVE} \Delta t \hspace{-100pt} &  \forall h, \forall t & \label{eq:soc_evs} \\
    & p_{h,t}^{\eta,v} = p_{h,t}^{v,\eta} = 0 \hspace{-100pt} & \hspace{-100pt} \forall t \notin [t^{DEP},t^{ARR}] & \label{eq:ev_connected} \\
    & \underline{SOC}_h \leq soc_{h,t} \leq \overline{SOC}_h & \forall h, \forall t & \label{eq:soc_limits_evs} \\
    & q_{h,t}^{HE} \leq Q_h^{HE}, \quad 0 \leq p_{h,t}^{CO} \leq P_h^{CO} & \forall h, \forall t & \label{eq:max_he_co_household} \\
    \nonumber & \tau_{h,t} = \tau_{h,t-1} + \frac{\Delta t}{R_h C_h}(\tau_{h,t}^{out} - \tau_{h,t-1}) + \frac{\Delta t}{C_h}(q_{h,t}^{HE} - p_{h,t}^{CO} \eta_{h}^{CO}) \hspace{-100pt} && \\ 
    & \hspace{100pt}  & \forall h, \forall t & \label{eq:tau_household} \\
    & \tau_h^{MIN} \leq \tau_{h,t} \leq \tau_h^{MAX} & \forall h, \forall t & \label{eq:tau_limit_households} \\
    & p_t^{s,e} \leq z_t^{BAT} P^{BAT}, \quad p_t^{e,s} \leq (1-z_t^{BAT}) P^{BAT} \hspace{-100pt} & \forall t & \label{eq:power_ec_battery}\\
    & soc_{t} = soc_{t-1} + (p_{t}^{e,s} \eta^{CH} - p_{t}^{s,e}/\eta^{DIS}) \Delta t \hspace{-100pt}  & \forall t &\label{eq:soc_ec_battery} \\
    & \underline{SOC} \leq soc_t \leq \overline{SOC} & \forall t & \label{eq:limit_ec_battery} \\
    & q_t^{DtC}\eta^{TR} + p_t^A\eta^A = \sum_h q_{h,t}^{HE} + q_t^{EXT} & \forall t & \label{eq:heat_transfer} \\
    & \eqref{eq:power_datacentre} - \eqref{eq:heat_data centre} & \label{eq:dtc_ems}
    \end{flalign}
    \label{eq:ems_ec}
\end{subequations}

Equation \eqref{eq:ems_balance} defines the energy balance of the \gls{eoa} module, while \eqref{eq:household_balance} computes the balance for each household. 
The household power consumption $p_{h,t}^{\eta,f}$ is modelled as a flexible demand with lower and upper bounds $\underline{P}_{h,t} \leq p_{h,t}^{\eta,f} \leq \overline{P}_{h,t}$, which comprises a baseline $P_{h,t}^{sch}$ modified in upward $p_{h,t}^{u}$ and downward $p_{h,t}^{d}$ directions by \eqref{eq:ec_fl_consumption}. 
The total consumption throughout the optimization horizon is enforced by \eqref{eq:ec_fl_total}. 
The charging $p_{h,t}^{\eta,v}$ and discharging $p_{h,t}^{v,\eta}$ power of the \gls{ev} is set to zero by \eqref{eq:ev_connected} if $t$ is lower than the arrival time $t^{ARR}$ or higher than the departure time $t^{DEP}$.
Considering $P_h^{EV}$ as the rating of the \gls{ev} charger, simultaneous charging and discharging is restricted by \eqref{eq:ev_ch_dis} using binary variable $z_{h,t}^{ev}$.
The \gls{soc} of each household's \gls{ev} depends on the charging and discharging powers and on the power needed for driving purposes $P_{h,t}^{DRIVE}$ by \eqref{eq:soc_evs}.
\gls{soc} limits $\underline{SOC}_{h}, \overline{SOC}_{h}$ are set by \eqref{eq:soc_limits_evs}.
Maximum heating $Q_h^{HE}$ and cooling transfer $P_h^{CO}$ for each household are in \eqref{eq:max_he_co_household}, while temperature $\tau_{h,t}$ constraints are defined in \eqref{eq:tau_household} and \eqref{eq:tau_limit_households} based on the thermal resistances $C_h$ and $R_h$ and outdoors temperature $\tau_{h,t}^{out}$ using a first-order linear model. 
Heating power $q_{h,t}^{HE}$ is obtained from the shared heating system, while $p_{h,t}^{CO}$ is used to drive cooling system with efficiency $\eta_h^{CO}$.
\gls{lec} shared battery is managed from \eqref{eq:power_ec_battery} to \eqref{eq:limit_ec_battery}, considering charging and discharging efficiencies $\eta^{CH}$ and $\eta^{DIS}$, a converter rating $P^{BAT}$, and \gls{soc} limits $\underline{SOC}$ and $\overline{SOC}$.
Simultaneous charging and discharging for \glspl{ev} and the shared \gls{lec} battery is prohibited by binaries $z_{h,t}^{ev}$ and $z_t^{BAT}$, respectively. 
The heat balance of the recovery system is described by \eqref{eq:heat_transfer}, for a household heating demand $\sum_h q_{h,t}^{HE}$. 
Lastly, \eqref{eq:dtc_ems} represents the \gls{dc} model.

\section{Case Study and Simulation Results}
In this section, a case study based on realistic \gls{lec} \cite{Goncalves2022} and \gls{dc} \cite{Wang2019a} information is presented. The workload \gls{dc} information is directly obtained from Oak Ride National Laboratory \cite{ornlLandingPage}.
The \gls{lec} has 10 households with consumptions ranging between 3 and 10 kW, photovoltaic panels of similar size and, one \gls{ev} per household. 
Each \gls{ev} typically travels between 50 and 100 kilometres per day, consuming 15 to 20 kWh per 100 km. 
The \gls{eoa} module optimise the overall energy costs as stated by \eqref{eq:ems_obj}.
The assessment evaluates the value of the synergies between \glspl{dc} and \glspl{lec}, also showing the impact of the number of households included in the system. 
Simulations were performed using Python 3.9.2 in PYOMO and solved by Gurobi 10.0.1 on an Apple M1 with 16 GB of RAM.
The assessment problem has 16,597 constraints, 2,887 variables and, 348 binaries for each day of simulation with hourly resolution.
The solving time for a one-day simulation is 9.92 s, while for a 15-day simulation it is 288.9 s.

\subsection{Synergies between LECs and DtCs}
\begin{figure}[b]
    \centering
    \includegraphics[width=\columnwidth]{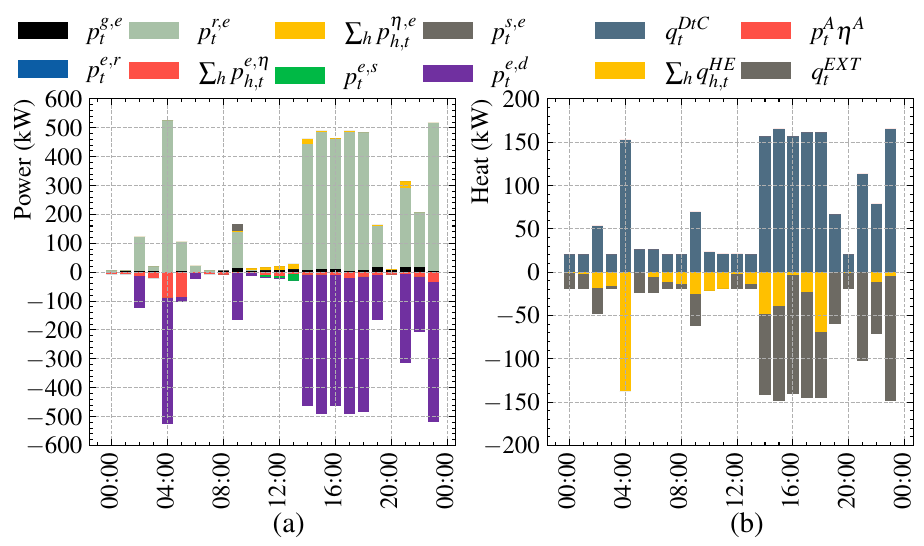}
    \caption{Energy (a) and Heat (b) balance of the EOA of the LEC. Negative values represent energy imports while positive values are exports.}
    \label{fig:ems_balance}
\end{figure}

This section showcases the synergies between \glspl{lec} and \glspl{dc} regarding waste heat and optimal market participation. 
Fig. \ref{fig:ems_balance} depicts the energy balance of the \gls{lec} for a typical operation day with low computational requirements. 
Still for an \gls{lec} of 10 households the \gls{dc} draw a significant portion of the energy, as presented in Fig. \ref{fig:ems_balance} (a) with purple bars. 
Nevertheless, this feature enables covering the heating needs of the households primarily based on the waste heat recovery. 
This is shown in Fig. \ref{fig:ems_balance} (b) with yellow bars, showing that there is no need of driving the \gls{hvac} $p_t^A$.

\begin{figure}
    \centering
    \includegraphics[width=\columnwidth]{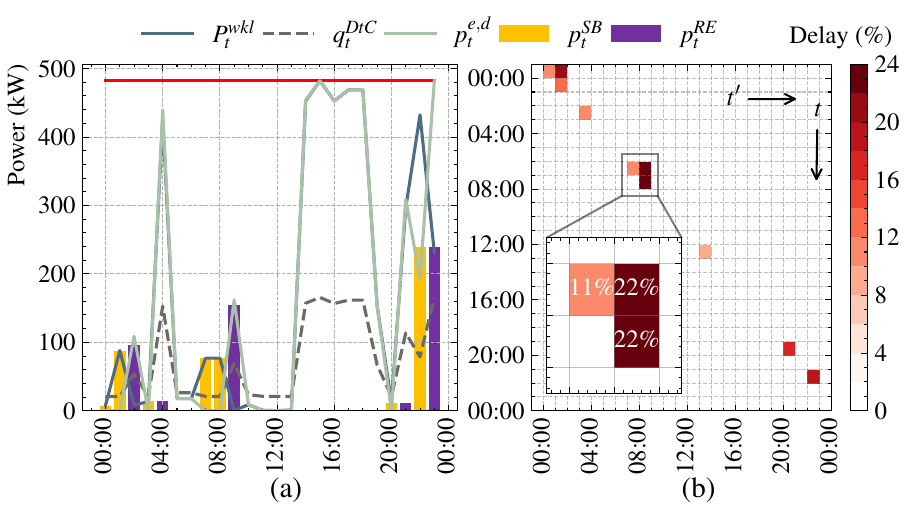}
    \caption{\gls{dc} Energy consumption (a) and delay matrix (b) for a day of operation with low computational requirements. Vertical axis represents the pausing time $t$, while the horizontal axis representing the re-starting time $t'$.}
    \label{fig:dctr_gen_work}
\end{figure}

A detailed view of the final consumption $p_t^{e,d}$ of the \gls{dc}, original workloads $P_t^{wkl}$, heat generated $q_t^{DtC}$ and pausing $p_t^{SB}$ and re-starting powers $p_t^{RE}$ is depicted in Fig. \ref{fig:dctr_gen_work} (a) for the day under study. 
As a result, the delay matrix for the preempted jobs is shown in Fig. \ref{fig:dctr_gen_work} (b), showing the increase in $d_{t,t'}$ when jobs are being paused in two consecutive periods between 7:00 and 9:00.
The delay for this case study was required to be less than 25\% of the original work duration. The vertical and horizontal axis represent pausing $t$ and re-starting $t'$ times while the colour indicates the magnitude of the delay $d_{t,t'}$. 
Given the duration of the workload $\mu_t$, and in order to meet \gls{qos} requirements, the paused power is resumed a maximum of two hours after pausing for this specific day.


\subsection{Long run assessment}
Using one year of data split into 24 simulations of 15 days each, the long-run performance of the coupling is evaluated in Table \ref{tab:long_run}.
The Thermal Coupling (TC) + Job Pausing (JP) + \gls{afrr} scenario showcases a 37.88\% reduction in operational costs compared to the No Coupling (NC) scenario. However, this cost-effectiveness does not extend proportionally to other metrics. For instance, the energy required from the retailer is only reduced by 5.81\%, 6.61\% and, 4.14\% if we compare NC scenario to TC, TC+JP and TC+JP+\gls{afrr}, respectively. 
Additionally, although renewable energy production remains constant across all scenarios, the utilization of this energy varies. The auto consumption ratio stays steady at around 35\%, and the additional heating requirements $p_t^{A}\eta^A$ were reduced by 6.98\%, 10.58\% and 7.85\% compared to the rest of the scenarios. 
These strategies minimise wastage and enhance heat reuse, aligning with sustainable energy practices, while the percentage of waste heat reuse is around 87\% for the last three scenarios. This showcase the potential of the \gls{dc} to provide heating to the households, reducing the need for additional heating sources. 
Nevertheless, the increase in heat recovery ratio is driven more by the job pausing strategy than market participation.

\begin{table}
\centering
\caption{Long run analysis. Scenarios are NC: No coupling, TC: Thermal Coupling, JP: Job Pausing and aFRR participation. }
\label{tab:long_run}
\begin{tabular}{lcccc}
\hline
 &
  \textbf{NC} &
  \textbf{TC} &
  \textbf{TC + JP} &
  \textbf{\begin{tabular}[c]{@{}c@{}}TC + JP \\ + aFRR\end{tabular}} \\ \hline
\textbf{Op. Costs (k€)} &
  95,136.47 &
  89,771.18 &
  88,849.72&
  59,103.19 \\
\textbf{\begin{tabular}[c]{@{}l@{}}Renewable\\ Gen. (MWh)\end{tabular}} &
  300.62 &
  300.62 &
  300.62 &
  300.62 \\
\textbf{\begin{tabular}[c]{@{}l@{}}Retailer's \\ Energy (MWh)\end{tabular}} &
  576.52 &
  543.03 &
  543.17 &
  552.67 \\
\textbf{\begin{tabular}[c]{@{}l@{}}AC Ratio (\%)\end{tabular}} &
  34.27\% &
  35.63\% &
  35.63\% &
  35.23\% \\
\textbf{\begin{tabular}[c]{@{}l@{}}Data Centre RE \\ Consumption (\%)\end{tabular}} &
  33.30\% &
  28.21\% &
  27.71\% &
  27.97\% \\
\textbf{\begin{tabular}[c]{@{}l@{}}Average \\ Job Delay (\%)\end{tabular}} &
  0.00\% &
  0.00\% &
  18.62\%&
  15.86\% \\
\textbf{\begin{tabular}[c]{@{}l@{}}Heating \\ (MWh)\end{tabular}} &
  118.69 &
  110.40 &
  106.13&
  109.37 \\
\textbf{\begin{tabular}[c]{@{}l@{}}Power to \\ \gls{hvac} (MWh)\end{tabular}} &
  118.69 &
  15.79 &
  15.67&
  15.76 \\
\textbf{\begin{tabular}[c]{@{}l@{}}Heat recovery \\ ratio (\%)\end{tabular}} &
  0.00\% &
  87.49\% &
  87.13\%&
  87.41\% \\ \hline
\end{tabular}
\end{table}

\subsection{Impact of the number of households}
\begin{figure}[ht]
    \centering
    \includegraphics[width=\columnwidth]{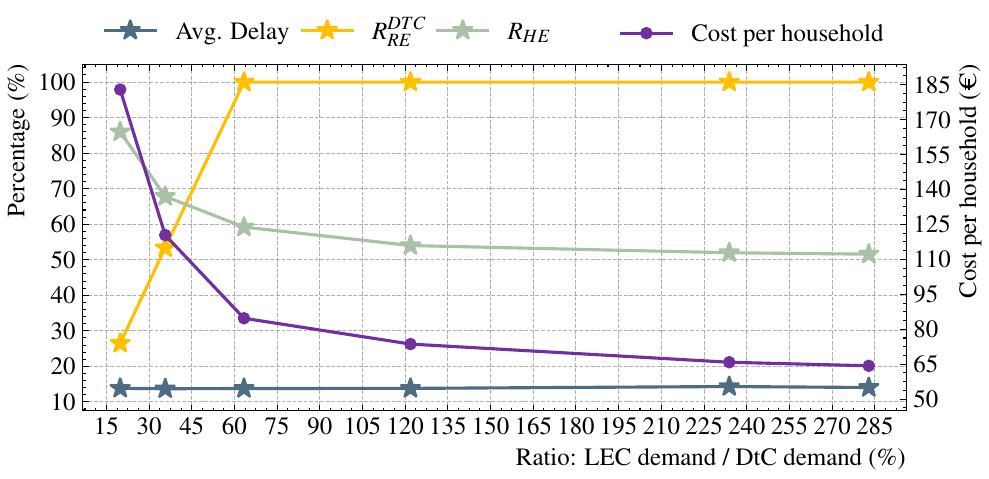}
    \caption{Impact of the ratio of the EC over the DtC demand on the \gls{qos} (blue), the percentage of renewable usage in the \gls{dc} (yellow), the ratio of heat recovery (green) and the cost per household (purple). From left to right, the number of households are 5, 10, 20, 40, 80, 100.}
    \label{fig:scalability}
\end{figure}

A sensitivity analysis has been performed to examine the impact of varying the number of households over a span of 10 days. 
Fig. \ref{fig:scalability} shows the impact of the number of connected households on the average job delay, the percentage of renewable used by \gls{dc} $R_{RE}^{DTC}$, the ratio of heat recovered by households $R_{HE}$ and the operational cost per household.
This in investigated through the ratio of the \gls{lec} net demand $\sum_{h,t} (p_{h,t}^{e,\eta}-p_{h,t}^{\eta,e})$ over the \gls{dc} demand $\sum_t p_t^{e,d}$.
The \gls{qos} remains stable and under the maximum requirement $K^{DELAY}$ for this case study regardless of the number of households introduced into the \gls{lec}. 
Furthermore, the \gls{dc} can meet its energy demands by utilising surplus energy from households equivalent of its 60\% demand. 
It is important to clarify that this does not indicate an exclusive reliance on renewable sources, as market trading with the retailer can occur. 
The cost per household is reduced as their number increases, showing a 42.85\% reduction compared to the case with 5 households.
However, the ratio $R_{HE}$ also decreases with the number of households, due to the increasing needs for heating, which can be also covered by renewable surplus of each household.

\section{Conclusion}
This paper undertakes an assessment of the value of the synergies among \glspl{lec}, \glspl{dc}, and energy markets, exploring the potential savings regarding energy cost, waste heat reuse and job scheduling flexibility.
A \gls{milp} model is used to evaluate the joint performance of \glspl{lec}, \glspl{dc}, and energy markets, using the Big-M method to model the \gls{qos} requirement of the \gls{dc}. 
The case study results show substantial benefits, including potential reductions in operating costs, up to 38\% reduction in heat demand when these synergies are exploited. 
Introducing job pausing contribute to better adapt to renewable variability, but it will reduce \gls{dc}'s \gls{qos}.
The trade-off between \gls{qos} and energy system benefits must be carefully considered, since the \gls{dc} flexibility obtains more benefits if used for market participation.
In addition, up to 43\% reduction in individualised operational costs can be obtained if the size of the \gls{lec} increases from 5 to 100 households. 
These findings underscore the potential of combining \glspl{lec} and \glspl{dc} to address modern energy management challenges. Future work will investigate possible formulations of online energy management systems for joint optimisation of \glspl{dc} and \glspl{lec}.

\appendices
\section{Reformulation of job delay computation}
The job delay $d_{t,t'}$ is computed as $(t'-t)/ \mu_t$ if and only if $\sum_{\tau = 0}^{t'} P_{\tau}^{RE} \leq \sum_{\tau = 0}^{t} P_{\tau}^{SB}$, else $d_{t,t'} = 0$. This restriction cannot be directly solved by commercial solvers. In order to do so, the Big-M method is applied, and this condition is reformulated as follows,
\begin{subequations}
\begin{flalign}
    & \sum_{\tau = 0}^{t'} P_{\tau}^{RE} - \sum_{\tau = 0}^{t} P_{\tau}^{SB} \leq  M(1-\delta_{t,t'}) & \forall t, \forall t'>t& \\
    & \sum_{\tau = 0}^{t'} P_{\tau}^{RE} - \sum_{\tau = 0}^{t} P_{\tau}^{SB} \geq - M\delta_{t,t'} & \forall t, \forall t'>t & \\
    & d_{t,t'}\delta_{t,t'} \leq K^{DELAY} & \forall t, \forall t'>t & 
\end{flalign}
\end{subequations}
with $M$ being a large enough parameter, set to $10^4$ for this case, and $\delta_{t,t'}$ being a binary variable indicating the way $d_{t,t'}$ is computed.

\ifCLASSOPTIONcaptionsoff
  \newpage
\fi

\bibliographystyle{IEEEtran}
\bibliography{references.bib}

\begin{thebibliography}{10}
\providecommand{\url}[1]{#1}
\csname url@samestyle\endcsname
\providecommand{\newblock}{\relax}
\providecommand{\bibinfo}[2]{#2}
\providecommand{\BIBentrySTDinterwordspacing}{\spaceskip=0pt\relax}
\providecommand{\BIBentryALTinterwordstretchfactor}{4}
\providecommand{\BIBentryALTinterwordspacing}{\spaceskip=\fontdimen2\font plus
\BIBentryALTinterwordstretchfactor\fontdimen3\font minus \fontdimen4\font\relax}
\providecommand{\BIBforeignlanguage}[2]{{%
\expandafter\ifx\csname l@#1\endcsname\relax
\typeout{** WARNING: IEEEtran.bst: No hyphenation pattern has been}%
\typeout{** loaded for the language `#1'. Using the pattern for}%
\typeout{** the default language instead.}%
\else
\language=\csname l@#1\endcsname
\fi
#2}}
\providecommand{\BIBdecl}{\relax}
\BIBdecl

\bibitem{F.G.Reis2021}
I.~F.G.~Reis, I.~Gon{\c{c}}alves, M.~A.R.~Lopes, and C.~Henggeler~Antunes, ``{Business models for energy communities: A review of key issues and trends},'' \emph{Renewable and Sustainable Energy Reviews}, vol. 144, no. April, 2021.

\bibitem{Guo2021}
\BIBentryALTinterwordspacing
C.~Guo, F.~Luo, Z.~Cai, and Z.~Y. Dong, ``{Integrated energy systems of data centers and smart grids: State-of-the-art and future opportunities},'' \emph{Applied Energy}, vol. 301, no. July, p. 117474, 2021. [Online]. Available: \url{https://doi.org/10.1016/j.apenergy.2021.117474}
\BIBentrySTDinterwordspacing

\bibitem{Wang2019a}
F.~Wang, S.~Oral, S.~Sen, and N.~Imam, ``{Learning from Five-year Resource-Utilization Data of Titan System},'' in \emph{2019 IEEE International Conference on Cluster Computing}.\hskip 1em plus 0.5em minus 0.4em\relax IEEE, 9 2019, pp. 1--6.

\bibitem{Yin2023}
X.~Yin, C.~Ye, Y.~Ding, and Y.~Song, ``{Exploiting Internet Data Centers as Energy Prosumers in Integrated Electricity-Heat System},'' \emph{IEEE Transactions on Smart Grid}, vol.~14, no.~1, pp. 167--182, 2023.

\bibitem{Garimella2013}
\BIBentryALTinterwordspacing
S.~V. Garimella, T.~Persoons, J.~Weibel, and L.~T. Yeh, ``{Technological drivers in data centers and telecom systems: Multiscale thermal, electrical, and energy management},'' \emph{Applied Energy}, vol. 107, pp. 66--80, 2013. [Online]. Available: \url{http://dx.doi.org/10.1016/j.apenergy.2013.02.047}
\BIBentrySTDinterwordspacing

\bibitem{ZacharyWoodruff2014}
J.~Zachary~Woodruff, P.~Brenner, A.~P. Buccellato, and D.~B. Go, ``{Environmentally opportunistic computing: A distributed waste heat reutilization approach to energy-efficient buildings and data centers},'' \emph{Energy and Buildings}, vol.~69, pp. 41--50, 2014.

\bibitem{Borland2023}
P.~L. Borland, K.~McDonnell, and M.~Harty, ``{Assessment of the Potential to Use the Expelled Heat Energy from a Typical Data Centre in Ireland for Alternative Farming Methods},'' \emph{Energies}, vol.~16, no.~18, p. 6704, 2023.

\bibitem{Fouladvand2022}
\BIBentryALTinterwordspacing
J.~Fouladvand, A.~Ghorbani, N.~Mouter, and P.~Herder, ``{Analysing community-based initiatives for heating and cooling: A systematic and critical review},'' \emph{Energy Research and Social Science}, vol.~88, no. November 2021, p. 102507, 2022. [Online]. Available: \url{https://doi.org/10.1016/j.erss.2022.102507}
\BIBentrySTDinterwordspacing

\bibitem{Huang2020}
\BIBentryALTinterwordspacing
P.~Huang, B.~Copertaro, X.~Zhang, J.~Shen, I.~L{\"{o}}fgren, M.~R{\"{o}}nnelid, J.~Fahlen, D.~Andersson, and M.~Svanfeldt, ``{A review of data centers as prosumers in district energy systems: Renewable energy integration and waste heat reuse for district heating},'' \emph{Applied Energy}, vol. 258, no. July 2019, p. 114109, 2020. [Online]. Available: \url{https://doi.org/10.1016/j.apenergy.2019.114109}
\BIBentrySTDinterwordspacing

\bibitem{Srithapon2023}
\BIBentryALTinterwordspacing
C.~Srithapon and D.~M{\aa}nsson, ``{Predictive control and coordination for energy community flexibility with electric vehicles, heat pumps and thermal energy storage},'' \emph{Applied Energy}, vol. 347, no. December 2022, p. 121500, 2023. [Online]. Available: \url{https://doi.org/10.1016/j.apenergy.2023.121500}
\BIBentrySTDinterwordspacing

\bibitem{Backe2021}
S.~Backe, M.~Korp{\aa}s, and A.~Tomasgard, ``{Heat and electric vehicle flexibility in the European power system: A case study of Norwegian energy communities},'' \emph{International Journal of Electrical Power and Energy Systems}, vol. 125, no. June 2020, p. 106479, 2021.

\bibitem{Zhang2022}
Y.~Zhang, D.~C. Wilson, I.~C. Paschalidis, and A.~K. Coskun, ``{HPC Data Center Participation in Demand Response: An Adaptive Policy with QoS Assurance},'' \emph{IEEE Transactions on Sustainable Computing}, vol.~7, no.~1, pp. 157--171, 2022.

\bibitem{Jahanshahi2022}
A.~Jahanshahi, N.~Yu, and D.~Wong, ``{PowerMorph: QoS-Aware Server Power Reshaping for Data Center Regulation Service},'' \emph{ACM Transactions on Architecture and Code Optimization}, vol.~19, no.~3, 2022.

\bibitem{Verma2015}
A.~Verma, L.~Pedrosa, M.~Korupolu, D.~Oppenheimer, E.~Tune, and J.~Wilkes, ``{Large-scale cluster management at Google with Borg},'' \emph{Proceedings of the 10th European Conference on Computer Systems, EuroSys 2015}, 2015.

\bibitem{Goncalves2022}
C.~Goncalves, R.~Barreto, P.~Faria, L.~Gomes, and Z.~Vale, ``{Dataset of an energy community's generation and consumption with appliance allocation},'' \emph{Data in Brief}, vol.~45, p. 108590, 2022.

\bibitem{ornlLandingPage}
``{SMC 2021 Data Challenge: Analyzing Resource Utilization and User Behavior on Titan Supercomputer},'' \url{https://doi.ccs.ornl.gov/ui/doi/334}, [Accessed 13-09-2024].

\end{thebibliography}

\end{document}